\documentclass{sig-alternate-2013}

\permission{Permission to make digital or hard copies of part or all of this work for personal or classroom use is granted without fee provided that copies are not made or distributed for profit or commercial advantage, and that copies bear this notice and the full citation on the first page. Copyrights for third-party components of this work must be honored. For all other uses, contact the owner/author(s). Copyright is held by the author/owner(s).}
\conferenceinfo{WWW 2015 Companion,}{May 18--22, 2015, Florence, Italy.}
\copyrightetc{ACM \the\acmcopyr}
\crdata{978-1-4503-3473-0/15/05. \\
http://dx.doi.org/10.1145/2740908.2742744}

\begin{document}
\conferenceinfo{WWW}{ 2015, May 18-22, 2015, Florence, Italy}

\title{Modeling and Predicting Popularity Dynamics of Microblogs using Self-Excited Hawkes Processes}

\numberofauthors{3}
\author{
	Peng Bao, Hua-Wei Shen, Xiaolong Jin, Xue-Qi Cheng\\
	\affaddr{Institute of Computing Technology, Chinese Academy of Sciences, Beijing, China}\\
	\email{pengbaocn@gmail.com, \{shenhuawei, jinxiaolong, cxq\}@ict.ac.cn}
}

\maketitle
\begin{abstract}
The ability to model and predict the popularity dynamics of \textit{individual} user generated items on online media has important implications in a wide range of areas. In this paper, we propose a probabilistic model using a Self-Excited Hawkes Process~(SEHP) to characterize the process through which individual microblogs gain their popularity. This model explicitly captures the triggering effect of each forwarding, distinguishing itself from the reinforced Poisson process based model where all previous forwardings are simply aggregated as a single triggering effect. We validate the proposed model by applying it on Sina Weibo, the most popular microblogging network in China. Experimental results demonstrate that the SEHP model consistently outperforms the model based on reinforced Poisson process.
\end{abstract}

\category{J.4}{SOCIAL AND BEHAVIORAL SCIENCES}{Sociology}
\category{H.4}{INFORMATION SYSTEMS APPLICATIONS}{Miscellaneous}

\terms{Measurement; Experimentation}

\keywords{popularity prediction; popularity dynamics; microblogs}

\section{Introduction}

With the explosive growth of User Generated Contents (UGC) on online media, it becomes an important issue to predict the popularity dynamics of UGC items, including microblogs, tweets, videos, to name a few. Popularity prediction has important implications in many domains, including viral marketing, public opinion monitoring, etc. Early studies devote to characterizing the distribution of the popularity over an aggregation of UGC items~\cite{Crane2008PNAS} and making prediction by exploiting temporal correlations~\cite{Bao2013WWW, Szabo2010CACM}.

Recently, researchers began to model the popularity dynamics of individual UGC items~\cite{Gomez2013ICML, Shen2014AAAI}. However, these models usually assume an aggregate stochastic process without distinguishing the triggering effects of different forwarding actions in the diffusion-and-reaction process. Therefore, although these models gain success in predicting, say, the citation counts of scientific papers and view counts of Youtube videos, they are not applicable to model popularity dynamics over a microblogging network, where interactions among users matter much in popularity dynamics.

In this paper, we propose a probabilistic model using a Self-Excited Hawkes Process~(SEHP) to model the process through which individual microblogs gain their popularity. This model explicitly captures the triggering effect of each forwarding, distinguishing itself from the reinforced Poisson process~(RPP) based model presented in~\cite{Shen2014AAAI}, where all previous forwardings are simply aggregated as a single triggering effect~(see Fig. 1). We validate the proposed model by applying it on Sina Weibo\footnote{http://t.cn}, the most popular microblogging network in China. Experimental results demonstrate that this model consistently outperforms the model based on reinforced Poisson process.

\section{The SEHP Model}

When a microblog spreads, it creates a \textit{cascade} on the microblogging network. The popularity dynamics of each microblog during observed time period $[0,T]$ can be characterized by a set of time stamps ${t_i}~(1\leq{i}\leq{N})$ which denote the occurrence time of each forwardings. Here, $N$ is the total number of forwardings. Without loss of generality, we have $0 = t_0 \leq{t_1} \leq{t_2} \leq{. . .} \leq{t_i} \leq{. . .} \leq{t_N} \leq{T}$. For a microblog, we model its popularity dynamics using an SEHP characterized with the following rate function
\begin{equation}\label{rateFunction}
\lambda(t) = v e^{-\beta t} + \alpha\!\sum_{j=1}^{j_{max}(t)}\!\!e^{-\beta (t-t_j)},
\end{equation}
where $v$ is the initial triggering strength that reflects the attractiveness of the microblog, $\alpha$ is the triggering strength of each subsequent forwarding, and $j_{max}(t)$ is the index of the last forwarding before time $t$. We set an exponential decaying function with exponent $\beta$ for simplicity.

\begin{figure}[!h]
\vspace{-5pt}
\centering
\includegraphics[height=1.5in, width=3.5in]{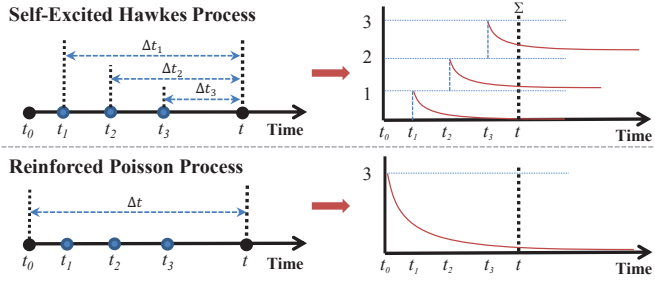}
\caption{Differences between SEHP and RPP}
\label{fig:structural}
\vspace{-5pt}
\end{figure}

According to the survival theory, given that the $(i-1)$-th forwarding arrives at $t_{i-1}$, the probability that the $i$-th forwarding arrives at $t_{i}$ follows
\begin{equation}\label{p1}
p(t_{i} | t_{i-1}) = e^{-\int_{t_{i-1}}^{t_{i}} \lambda(t)dt} \lambda(t_i),
\end{equation}
and the probability that no forwarding arrives between $t_N$ and $T$ is
\begin{equation}\label{p2}
p(T | t_N) = e^{-\int_{t_N}^{T} \lambda(t)dt}.
\end{equation}

Assuming that forwardings during different time intervals are statistically independent, the likelihood of observing a cascade of a microblog and its subsequent forwardings during time interval $[0,T]$ follows
\begin{equation}\label{eqlikelihood}
\begin{split}
\mathcal{L}(\alpha, \beta, v)
&= p(T | t_N)\prod_{i=1}^{N}{p(t_i | t_{i-1})}.
\end{split}
\end{equation}

By substituting Eqs. (\ref{rateFunction}), (\ref{p1}), and (\ref{p2}) in Eq. (\ref{eqlikelihood}), we obtain the logarithmic likelihood
\begin{equation}\label{eqloglikelihood}
\begin{split}
\log\mathcal{L}(\alpha, \beta, v) =
& \frac{v}{\beta}\left(e^{-\beta T}\!-\!1\right)\!+\!\frac{\alpha}{ \beta}\sum_{i=1}^{N}\left(e^{-\beta (T-t_i)}\!-\!1\right)\!+\!\\
&\sum_{i=1}^{N}\log\!\left(v e^{-\beta t_i}\!+\!\alpha\sum_{j=1}^{j_{max}(t_i)}\!\!e^{-\beta (t_i-t_j)}\right)\!\!.\end{split}
\end{equation}

We employ maximum likelihood estimation to infer the parameters in the proposed model. With the estimated parameters, the model can be used to predict the expected number $c(t)$ of forwardings of a microblog up to any given time $t$. With the rate function in Eq.~(\ref{rateFunction}), we obtain the prediction function
\begin{equation}
c(t)\!=\!N\!+\!\frac{v}{\beta}\!\!\left(\!e^{-\beta T}\!\!-\!e^{-\beta t}\!\right)\!+\!\frac{\alpha}{\beta}\!\!\!\sum_{j=1}^{j_{max}(t)}\!\!\!\!\left(\!e^{-\beta (T-t_j)}\!\!-\!e^{-\beta (t-t_j)}\!\right)\!\!.\!\!
\end{equation}

\section{Experimental Validation}
Experiments are conducted on a dataset from Sina Weibo, published by the WISE 2012
Challenge\footnote{http://www.wise2012.cs.ucy.ac.cy/challenge.html}. We select microblogs that were submitted during July 1-31, 2011 and have more than 10 forwardings during the first hour and more than 100 forwardings during forty-eight hours after submission. This resulting dataset consists of 5670 microblogs and their cascades.

To validate the prediction performance of the SEHP, we compare it with the state-of-the-art model based on reinforced Poisson process~\cite{Shen2014AAAI}, in terms of two metrics:
\begin{figure}[!h]
\vspace{-5pt}
\centering
	\includegraphics[height=1.2in, width=1.6in]{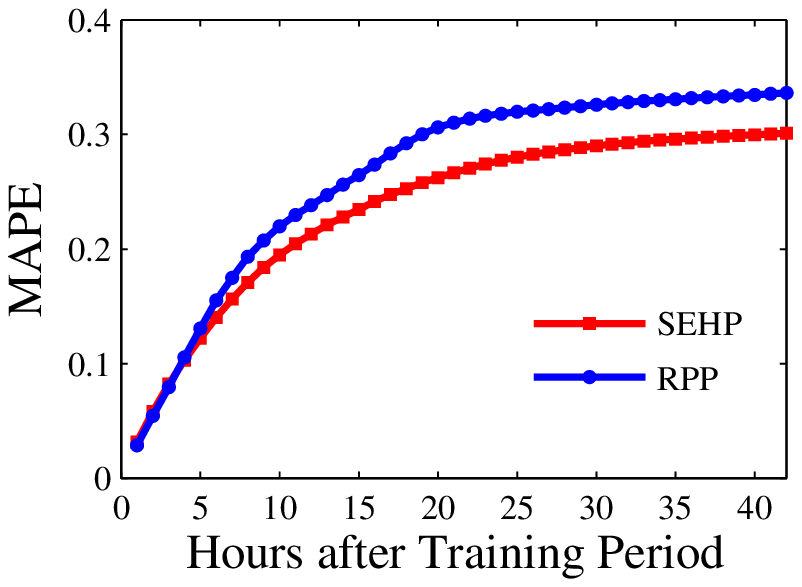}
	\includegraphics[height=1.2in, width=1.6in]{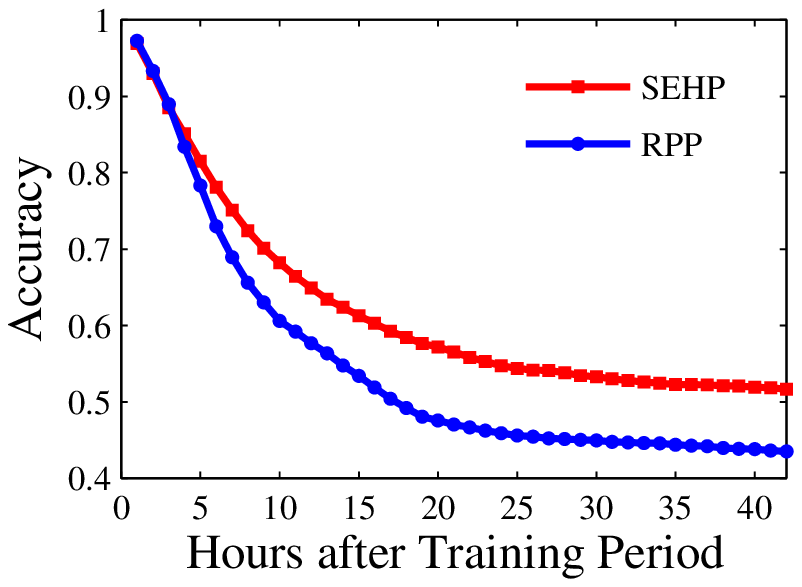}
\caption{Prediction performance}
\label{fig:structural}
\vspace{-5pt}
\end{figure}

\begin{itemize}
\item Mean Absolute Percentage Error~($MAPE$): It measures the average derivation between the predicted and observed popularity over all microblogs. Denoting the predicted popularity for a microblog $i$ up to time $t$ as $c_i (t)$ and its actual popularity as $r_i (t)$, the MAPE over $M$ microblogs can be written as
\begin{displaymath}
MAPE = \frac{1}{M} \sum_{i=1}^{M} \left| \frac{c_i (t)- r_i (t)}{r_i (t)} \right|.
\end{displaymath}
\item $Accuracy$: It measures the fraction of microblogs, correctly predicted under a given error tolerance $\epsilon$. Specifically, the accuracy of popularity prediction over $M$ microblogs  is
\begin{displaymath}
Accuracy = \frac{1}{M} \left| \left\{i: \left| \frac{c_i (t)- r_i (t)}{r_i (t)} \right| \leq \epsilon \right\} \right|.
\end{displaymath}
The threshold $\epsilon$ is set as 0.2 in this paper.
\end{itemize}

We set the training period, i.e., $T$, as 6 hours and then predict the popularity for each microblog from the 1st to 42nd hour after the training period. As shown in Fig. 2, the SEHP model consistently exhibits lower error and higher accuracy than the RPP model.

\section{Conclusions}
In this paper, we proposed a probabilistic model to characterize and predict the popularity dynamics of microblogs using an SEHP. Experiments on a Sina Weibo dataset demonstrated that this model consistently outperforms the baseline model based on reinforced Poisson process.

\section{Acknowledgements}
This work is funded by the 973 Program of China (Nos. 2014CB340401 and 2012CB316303) and the NSFC (Nos. 61472400, 61425016, 61232010, 61272353, and 61370128). The authors would like to thank the NASC Research Group for valuable discussions and suggestions.


\begin{thebibliography}{6}

\bibitem{Bao2013WWW}
P. Bao, H.~W. Shen, J. Huang, X.~Q. Cheng. Popularity Prediction in Microblogging Network: a Case Study on Sina Weibo. In \textit{Proc. of  WWW '13}, pp. 177-178, Brazil.

\bibitem{Crane2008PNAS}
R. Crane, D. Sornette. Robust dynamic classes revealed by measuring the response function of a social system. \textit{Proc. Natl. Acad. Sci.}, 105(41): 15649-15653, 2008.


\bibitem{Gomez2013ICML}
M.~Gomez-Rodriguez, J.~Leskovec, B.~Sch$\ddot{o}$lkopf. Modeling Information Propagation with Survival Theory. In \textit{Proc. of ICML '13}, pp. 666--674, USA.

\bibitem{Shen2014AAAI}
H.~W. Shen, D. Wang, C. Song, A.-L. Barab\'asi. Modeling and Predicting Popularity Dynamics via Reinforced Poisson Processes. In \textit{Proc. of AAAI '14}, pp. 291-297, Canada.

\bibitem{Szabo2010CACM}
G.~Szabo, B.~A. Huberman. Predicting the popularity of online content. \textit{Commun. ACM}, 53(8): 80-88, 2010.


%
%
%
%
%
%
%

\end{thebibliography}
\end{document}